\documentclass[preprint,12pt]{elsarticle}

\usepackage{amssymb}
\usepackage{amsmath}
\usepackage{siunitx}

\journal{Journal of the Mechanics and Physics of Solids}

\begin{document}

\begin{frontmatter}



\title{Crack tip kinematics reveal the cohesive zone structure in brittle hydrogel fracture}


\author[inst1]{Chenzhuo Li}
\author[inst1]{Xinyue Wei}

\affiliation[inst1]{organization={Engineering Mechanics of Soft Interfaces, School of Engineering, Ecole Polytechnique Fédérale de Lausanne},
            city={Lausanne},
            postcode={1015}, 
            country={Switzerland}}

\author[inst2]{Meng Wang}

\affiliation[inst2]{organization={The Racah Institute of Physics, The Hebrew University of Jerusalem},
            city={Jerusalem},
            postcode={91904}, 
            country={Israel}}

\author[inst3]{Mokhtar Adda-Bedia}

\affiliation[inst3]{organization={Université de Lyon, Ecole Normale Supérieure de Lyon, CNRS, Laboratoire de Physique},
            city={Lyon},
            postcode={F-69342}, 
            country={France}}

\author[inst1]{John M. Kolinski}

\begin{abstract}
When brittle hydrogels fail, several mechanisms conspire to alter the state of stress near the tip of a crack, and it is challenging to identify which mechanism is dominant. In the fracture of brittle solids, a sufficient far-field stress results in the complete loss of structural strength as the material `unzips’ at the tip of a crack, where stresses are concentrated. Direct studies of the so-called small-scale yielding zone, where deformation is large, are sparing. Using hydrogels as a model brittle solid, we probe the small-scale yielding region with a combination of microscopy methods that resolve the kinematics of the deformation. A zone over which most of the energy is dissipated through the loss of cohesion is identified in the immediate surroundings of the crack tip. With direct measurements, we determine the scale and structure of this zone, and identify how the specific loss mechanisms in this hydrogel material might generalize for brittle material failure.
\end{abstract}

\begin{graphicalabstract}
\centering
\includegraphics[width=0.98\textwidth]{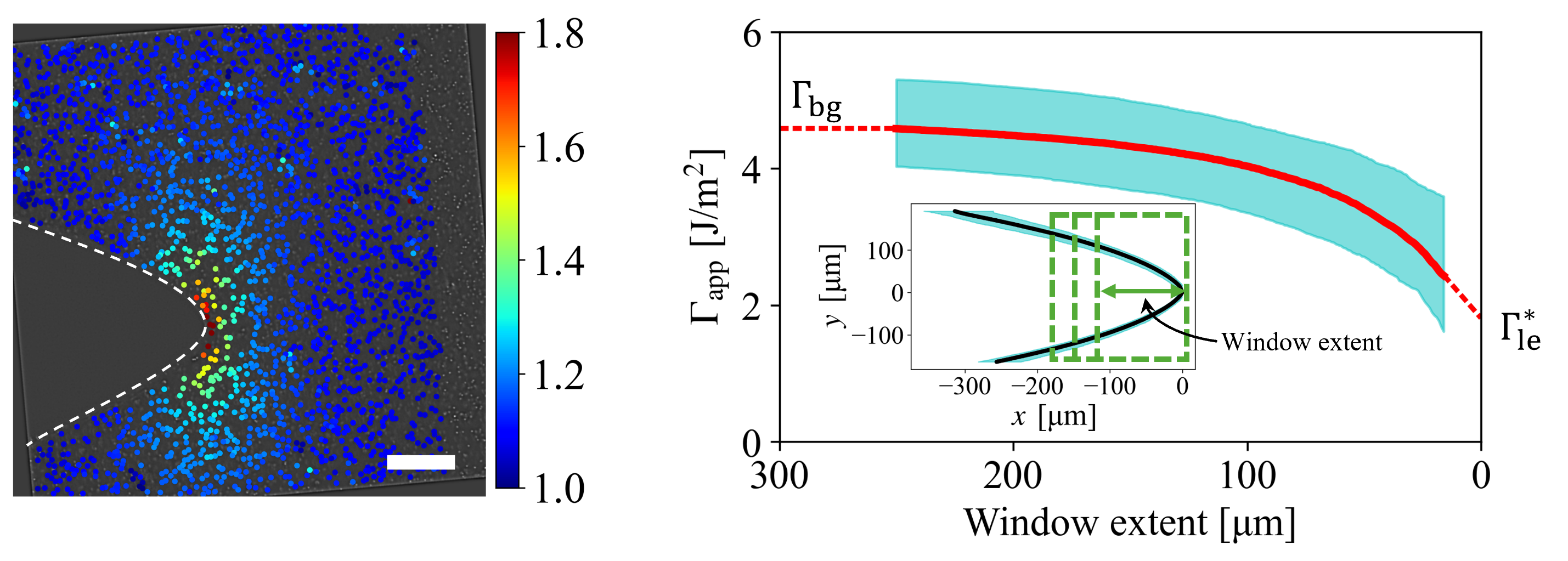}
\end{graphicalabstract}

\begin{highlights}
\item Brittle hydrogels fail cohesively
\item Energy dissipation near the crack tip occurs both locally at the crack tip, and in a distributed damage zone
\end{highlights}


\end{frontmatter}


\section{Introduction}
\label{sec:intro}
When brittle materials fail, they typically lose cohesion through the process of fracture~\cite{anderson}. In the existing theory of brittle fracture, the loss of cohesion is presumed to be confined to a small volume near the crack tip, as asserted by the `small-scale yielding' hypothesis~\cite{rice_1968, anderson}. Outside of this zone, the material is typically modelled as a linear elastic solid~\cite{williams_1952, rice_1968}, where the stress diverges near the crack tip with the characteristic `inverse-square-root’ $r^{-1/2}$ asymptotics. For soft materials such as hydrogels, the diverging stress is typically accompanied by large deformation prior to failure~\cite{goldman_boue_failing_2015, kolvin_how_2018} - thus, several candidate mechanisms that regulate the stress can become active, complicating the evaluation of the stress state near the crack tip. In the absence of a clear failure mechanism at the crack tip, significant open questions concerning the near crack tip fields remain unanswered~\cite{bouchbinder_1/r_2009, begley_2015}. Indeed, the large deformation in this region elicits constitutive non-linearity, poro-elastic solvent flux for hydrogels, and cohesive loss, rendering the nature of material failure at the crack tip ambiguous. Measurement of deformation in this region is further complicated by the singular nature of the crack tip field, which requires ever-increasing resolution to characterize the deformation field with data that can be used to identify the relevant failure mechanisms. Despite these challenges, direct measurements of deformation and kinematics within the small-scale region near the crack tip in a brittle solid can offer insight into how material failure occurs at the smallest scales, and thus illuminate the physics of brittle fracture generally.

To address the experimental challenges presented by the large deformation in the diverging stress field at the crack tip, we probe the material kinematics in the near-tip field of a quasi-static crack tip in a brittle polyacrylamide hydrogel under remote tensile loading. Brittle hydrogels are used as a proxy for the broad class of brittle solids, which have the merit of a lesser material sound speed, facile preparation, and an expansive application domain in their own right~\cite{tanaka_2000, livne_2005, baumberger2006fracture, livne_breakdown_2008, livne_2010, kolvin_topological_2017, zhao_chemrev}. Polyacrylamide hydrogels with a canonical composition that results in nearly ideally brittle material response were used (see \textit{Materials and Methods} for preparation protocol)~\cite{livne_breakdown_2008}. Two distinct, high-resolution optical methods are used in these experiments. In the first method, we embed fluorescent dye in the hydrogel, and monitor the crack tip opening displacement (CTOD) on a single plane extending a few hundred microns from the crack tip using confocal microscopy. In the second method, we embed particles in a hydrogel and measure the deformation within the hydrogel near the crack tip. The particles are sufficiently small that they do not alter the fracture dynamics~\cite{Taureg2020Particles}. These high-resolution measurements enable us to probe the deformation kinematics to within \SI{10}{\micro\meter} of the crack tip. 

Analysis of the near-tip kinematic data using known solutions for the crack tip opening displacement~\cite{Long_Hui_2015}, and direct evaluation of the $J$-integral~\cite{rice_1968} for neo-Hookean solids~\cite{knowles_1973, liu2020asymptotic}, show that the strain energy decreases strongly within approximately \SI{100}{\micro\meter} of the crack tip, indicating dissipation in a cohesive zone. The functional form of the cohesive law is evaluated from the decrease of the fracture energy near the crack tip. By windowing the boundary of the $J$-integral, the structure of the cohesive zone is measured. Careful characterization of the material, and established estimates for solvent fluxes~\cite{chester_2011, bouklas_2012, Long_Hui_2015, Baumberger_2020} are used to evaluate possible strain-stiffening and poro-elastic stress relaxation; we find that alternative mechanisms that could modify the material response near the crack tip are not active on the scale of the measured cohesive zone; instead, the mode of failure is consistent with a cohesive zone reminiscent of the Irwin-Orowon picture of network fracture~\cite{suo_1}. For dynamic cracks in the same hydrogel, a non-linear scale $\delta_{dyn}$ emerges near the crack tip ~\cite{livne_breakdown_2008}; this scale is attributed to the $1/r$ scaling that occurs at the crack tip of strain-stiffening materials described by a generalized neo-Hookean material model~\cite{knowles_1973}, or a weakly non-linear expansion of the displacement field\cite{bouchbinder_1/r_2009, goldman_boue_failing_2015}. While we observe a similar crack tip feature in our data, the requisite strain stiffening does not emerge at our strain rates, and thus a parabolic CTOD is expected for the nearly-ideal neo-Hookean material model that describes our hydrogel at the strain rates realized in our experiments~\cite{Long_Hui_2015}; thus, strain stiffening is not responsible for the sharper CTOD~\footnote{At higher strain rates, strain stiffening could emerge, but we cannot yet quantitatively analyze these data as the material response at the strain rates typically achieved in dynamic fracture remains to be characterized.}.  Finally, we discuss the implications of the value of the cohesive zone stress at the crack tip, and show that the small-scale but distributed damage in the cohesive zone surrounds a more localized process zone on a smaller scale than the resolution of our measurements, indicating a separation of scales between the cohesive zone and the process zone where the material ultimately separates. While the size of the cohesive zone should vary for different materials, the nature of brittle material failure identified here may be universal for fracture in brittle solids. 

\section{Materials and Methods}
\label{section:mat-methods}

\subsection{Confocal Microscopy for CTOD}
\label{subsection:mmctod}

\begin{figure}
\centering
\includegraphics[width=0.7\textwidth]{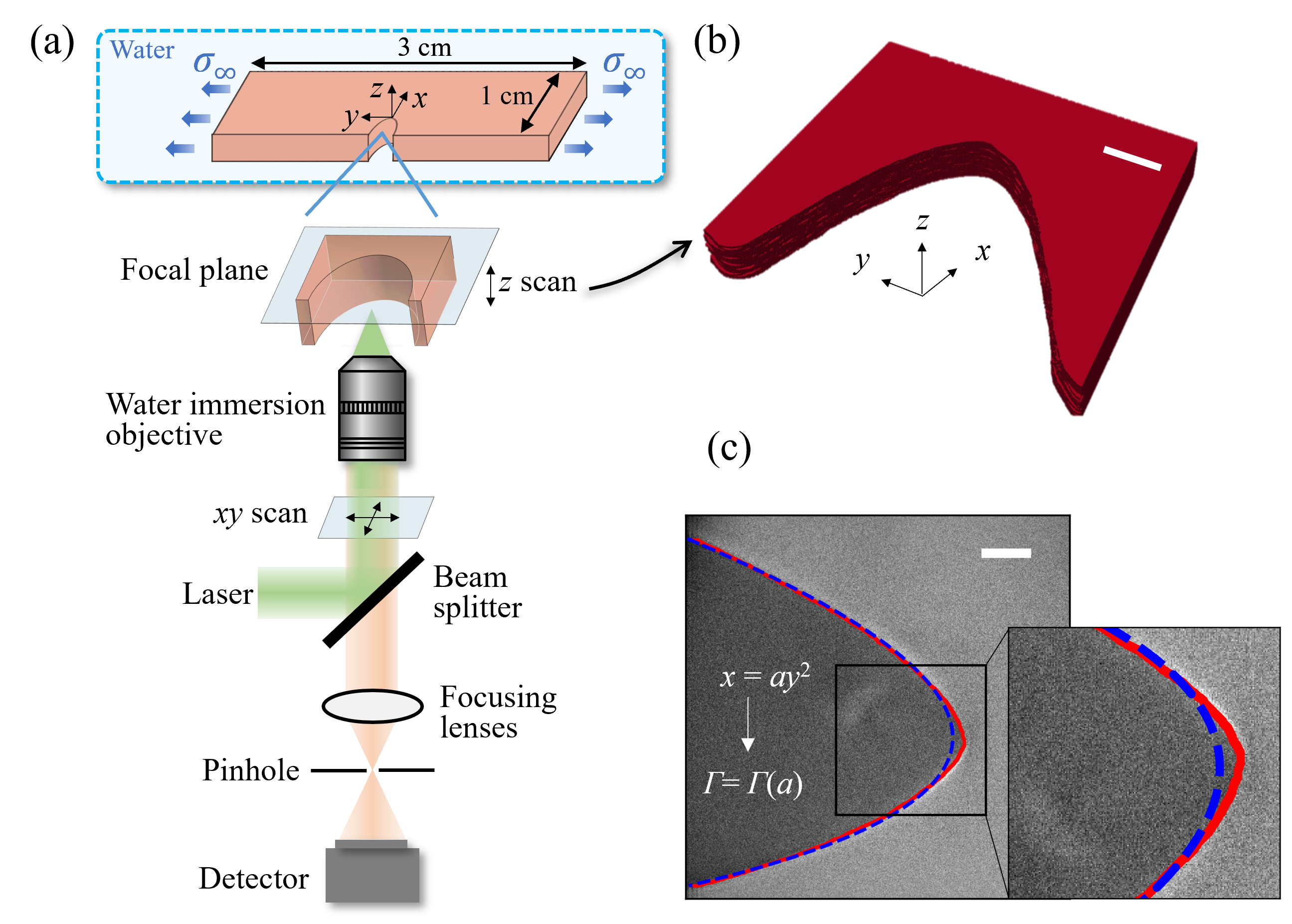}
\caption{Experimental setup for CTOD measurement near the crack tip with a confocal microscope.
(a) Schematic drawing of the confocal microscope system. The pre-cracked \SI{3}{\cm} $\times$ \SI{1}{\cm} $\times$ \SI{200}{\micro\meter} hydrogel sample is dyed by Rhodamine 6G, immersed in water, and subjected to mode~I loading via the sample grips. The laser beam (wavelength 525~nm) is focused into the sample by the objective (Nikon CFI Plan Fluor 10X W and CFI75 Apo 25XC W), exciting the florescent dye. The emission light passes through the dichroic beam splitter, and the out-of-focus light is blocked by the pinhole in front of the detector. By rastering the beam in $x$ and $y$, and scanning the objective in $z$, the confocal system collects the fully 3D geometry of the crack tip.
(b) A typical 3D reconstruction of a quasi-static planar crack, imaged with the $25\times$ objective. The crack speed is $\sim$\SI{2.5}{\micro\meter / \second}. The scale bar is \SI{100}{\micro\meter}.
(c) A 2D planar slice of the 3D reconstruction shown in (b). The scale bar is \SI{50}{\micro\meter}. The CTOD is determined with a custom image processing routine (using Python skimage packages~\cite{scikit2014}), and shown in the red solid curve. A parabola is fitted to the CTOD curve and plotted in the blue dashed line. The prefactor $a$ of the parabola $x=ay^2$ fitted to the CTOD data is related to the apparent fracture energy $\Gamma_\mathrm{app}$ using energy balance, as shown in Eq.~\ref{eq_CTOD}. At the crack tip, the fitted parabola deviates from the CTOD as shown in the zoomed-in view.}
\label{fig_CTOD_Setup}
\end{figure}

Experiments are performed on thin (\SI{200}{\micro\meter}) samples of polyacrylamide hydrogel whose solvent contains a fluorescent dye as described in Sec.~\ref{subsection:gel_preparation}. The experimental setup for CTOD measurement is schematically shown in Fig.~\ref{fig_CTOD_Setup}(a), where a confocal microscopy is used to capture the in-situ 3D information of the crack tip. The fluorescent sample with an edge crack is placed under far-field mode~I loading applied by displacement controlled stages. Using a pinhole to block the out-of-focus light, the detector only collects the emitted light from the focal point inside the sample. We constructed the 3D crack by taking a volume scan of the mode~I crack during quasi-static propagation ($\sim$\SI{10}{\micro\meter/\second}). A representative reconstructed 3D CTOD volume is shown in Fig.~\ref{fig_CTOD_Setup}(b). This 3D reconstruction demonstrates the planar symmetry of the crack, and ensures that the 2D CTOD is sufficiently representative of the fully 3D crack tip opening profile for the clean cracks~\cite{Wang2022JMPS} at the focus of the present work. 

\subsection{Microscopy of particle-laden hydrogel samples}
\label{subsection:mmparticles}

\begin{figure}[!ht]
\centering
\includegraphics[width=0.7\textwidth]{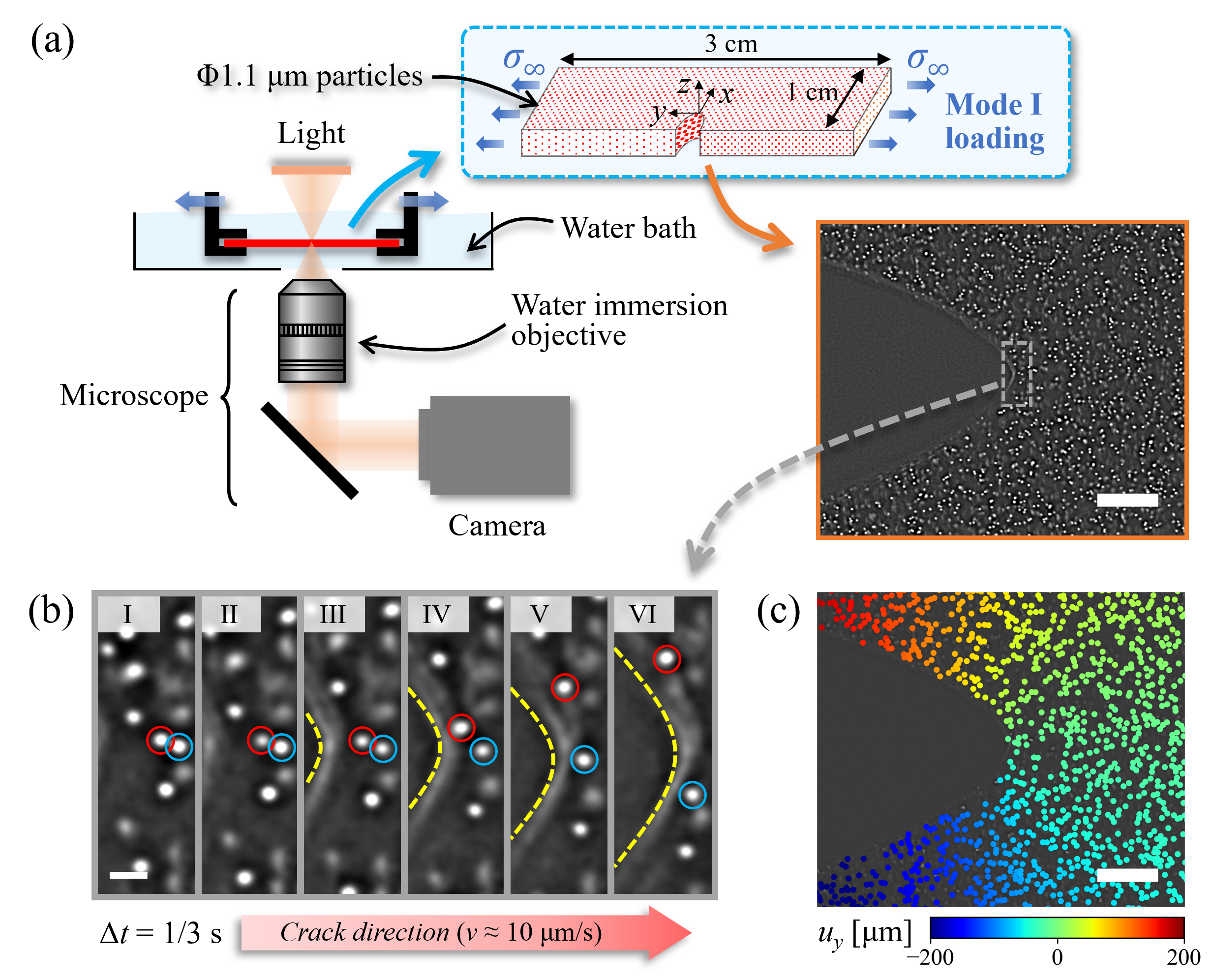}
\caption{
Experimental setup for full-field deformation measurement near the crack tip. (a) A thin (\SI{200}{\micro\meter}) sample of hydrogel containing embedded particles is gripped within a microscale tensile testing apparatus and immersed in a water bath. Before mounting, an edge crack is introduced centered between the grips, orthogonal to the loading axis. A Nikon TI eclipse microscope is configured for dark-field illumination with a $10\times$ water-immersion objective and images the focal plane onto the camera sensor. A sample image recorded with this configuration is pre-processed with bandpass filter and shown bottom right; the scale bar is \SI{100}{\micro\meter}.
(b) A time series obtained from the small region of interest indicated by the grey dashed line in (a) shows both the progress of the crack and the displacement of the embedded particles; two of approximately 4000 tracked particles are encircled with red and blue throughout the time series. The crack position is indicated in each frame by yellow dashed line. The scale bar shown in the first frame lower left is \SI{10}{\micro\meter}. The crack direction and speed is shown by the red arrow, bottom.
(c) After tracking each particle, its displacement from the initial position in the unstrained material is measured; the $y$ component of displacement, $u_y$, is indicated at each particle by color. In this way, material displacements are measured within \SI{10}{\micro\meter} of the crack tip. The scale bar is \SI{100}{\micro\meter}.
}
\label{fig_J_Setup}
\end{figure}

Experiments are furthermore performed on hydrogel samples embedded with particles, to confirm the near-tip energy dissipation by $J$-integral via full-field deformation measurement. The particle-embedded hydrogel was prepared as described in Sec.~\ref{subsection:gel_preparation} and mounted on the testing apparatus shown in Fig.~\ref{fig_J_Setup}(a). After carefully inserting an edge crack and loading the sample in tension, the crack begins to propagate quasi-statically, as indicated by the yellow dashed line in the time series in Fig.~\ref{fig_J_Setup}(b). Here, a small region of interest, particles in the vicinity of the crack tip are seen to remain bright, facilitating subsequent tracking within microns of the crack tip, which allows direct probing of the near-tip fields and distinguishes our analysis from the macroscale experiments \cite{Tracking_Macroscale}. Despite the large displacements and diverging path lines near the crack tip, the particle tracking algorithm remains robust. Example trajectories for two particles adjacent to the crack tip are highlighted by red and blue circles in the time series shown in Fig.~\ref{fig_J_Setup}(b). Once the particle tracking is complete and the reference locations are recovered by the procedure described in Fig.~\ref{figS_X0}, components of the displacement of each particle are directly recovered over the entire tracked region as demonstrated by the displacement component along the $y-$axis, $u_y$, in Fig.~\ref{fig_J_Setup}(c). 

\subsection{Preparation of hydrogel samples}\label{subsection:gel_preparation}
Two kinds of hydrogel samples, prepared with the same monomer chemistry, were used in our fracture experiments - hydrogel samples dyed with fluorescence dye and hydrogel samples embedded with micro particles.

To prepare the fluorescent samples, we used a precursor solution comprised of 13.8 wt.\% monomer and 2.7 wt.\% bis-acrylamide cross-linker, used in several prior brittle fracture studies\cite{livne_2005,livne_2010,goldman_acquisition_2010}. The stock solution is first degassed in a vacuum chamber for 10 minutes, and then 0.2 \% ammonium per-sulfate (APS) initiator and 0.02 \% tetramethylethylenediamine (TEMED) catalyst is added to trigger the free-radical polymerization reaction. After mixing for 30 seconds, the solution is poured into a \SI{190}{\micro\meter} gap between glass plates. The polymerization reaction runs for at least four hours. After polymerization, the top glass plate was removed, and the hydrogel was cut into \SI{3}{\centi\meter} by \SI{1}{\centi\meter} samples; these samples were then placed into $2 \times 10^{-4} \mathrm{mol/L}$ Rhodamine 6G solution to dye the solvent. After 24 hours immersed in the dye solution, the hydrogel samples were then ready to be mounted on the confocal microscope for the fracture experiments.

To prepare particle-embedded hydrogel, we mixed polystyrene particles (diameter \SI{1.1}{\micro\meter}) with the degassed monomer/cross-linker solution at a concentration 0.005 wt.\%, and sonicated for 5 minutes to disperse the aggregated particles and ensure their random distribution. The solution was was polymerized with APS and TEMED as described above, and poured into the same mold geometry. The polymerized hydrogel, now containing the polystyrene micro-particles, was then cut into samples of the same \SI{3}{\centi\meter} by \SI{1}{\centi\meter} size, and immersed in water for 24 hours to swell to an equilibrium state for fracture experiments in a waterbath.

As prepared, neither the Rhodamine 6G dye or the micro-particles are supposed not to affect the elastic properties or the fracture properties of the hydrogel. Therefore, the prepared hydrogels can be well modelled as incompressible neo-Hookean materials. The Young's modulus, fracture energy, and Rayleigh wave speed can be estimated as those of the pure hydrogel~\cite{livne_2005, Wang2022JMPS, goldman_acquisition_2010}, i.e., approximately \SI{90}{\kilo\pascal}, \SI{5.28}{\joule/\meter^2}, and \SI{5.5}{\meter/\second}, respectively.

\subsection{Derivation of apparent fracture energy and constitutive response from CTOD}
The evaluation of apparent fracture energy from the CTOD is based on linear elastic fracture mechanics (LEFM) theory. From LEFM, the mode~I displacement field~\cite{anderson} is
\begin{equation}\label{eq_LEFM_displacement}
    \begin{split}
        &u_x = \frac{K_I}{2\mu}\sqrt{\frac{r}{2\pi}} \cos{\frac{\theta}{2}} \left(\kappa - 1 + 2 \sin^2\frac{\theta}{2}\right),\\
        &u_y = \frac{K_I}{2\mu}\sqrt{\frac{r}{2\pi}} \sin{\frac{\theta}{2}} \left(\kappa + 1 - 2 \cos^2\frac{\theta}{2}\right),
    \end{split}
\end{equation}
where $r$ and $\theta$ are polar coordinates in the undeformed states, $\kappa = \frac{3-\nu}{1+\nu}$ for plane stress, and the Poisson's ratio $\nu = 0.5$ due to incompressibility. On the fracture surfaces $\theta = \pm \pi$, the displacement field in equation~\ref{eq_LEFM_displacement} becomes 
\begin{equation}\label{eq_uy}
    \begin{split}
        &u_x = 0,\\
        &u_y = \pm \frac{K_I}{2\mu}\sqrt{\frac{r}{2\pi}} (\kappa + 1).
    \end{split}
\end{equation}
Given $u_x = x-X$ and $u_y = y-Y$, the CTOD on the fracture surface is 
\begin{equation}
    x = -r = -\frac{9 \pi\mu^2 }{8 K_I^2} y^2 = -a y^2.
\end{equation}
The stress intensity factor $K_I$ is related to the strain energy release rate by $G = K_I^2/E=3K_I^2/\mu$ ($\mu=3E$ due to incompressibility); thus, $G$ can be derived from the prefactor $a$ of the parabolic CTOD predicted from LEFM. At the critically loaded state, this is equal to the fracture energy of the material by energy balance. Thus, the apparent fracture energy $\Gamma_{\text{app}}$ from LEFM is
\begin{equation}
    \Gamma_{\text{app}} = G_c = \frac{3 \pi \mu}{8 a}.
\end{equation}

The total measured fracture energy $\Gamma_\mathrm{app}$ consists of the energy dissipation in the cohesive zone $\Gamma_c$ and the constant energy dissipation $\Gamma^*$ at the crack tip, which is
\begin{equation}
    \Gamma_\mathrm{app} = \Gamma_\mathrm{c} + \Gamma^*.
\end{equation}
While $\Gamma^*$ remains constant, as it is localized at the crack, $\Gamma_c$ represents the work required per unit surface area in order to separate the crack faces to a separation distance $\delta_c$ by overcoming the cohesive stress $\sigma_c$; thus, $\Gamma_c$ is given by~\cite{freund}
\begin{equation}
    \Gamma_c = \int_0^{\delta_c} \sigma_c(\delta) d\delta.
\end{equation}

The constitutive response within the cohesive zone is thus given by
\begin{equation}
    \sigma_c = \frac{ d\Gamma_c}{d\delta}  = \frac{ d(\Gamma_\mathrm{app} - \Gamma^*)}{d\delta} = \frac{ d\Gamma_\mathrm{app}}{d\delta}.
\end{equation}

\subsection{Experimental measurement of deformation fields and calculation of $J$-integral}
We carried out the full-field deformation measurement near the crack tip using the experimental setup shown in Fig.~\ref{fig_J_Setup} (a). The hydrogel was dispersed with \SI{1.1}{\micro\meter} diameter polystyrene particles and was cut into \SI{3}{\centi\meter} by \SI{1}{\centi\meter} rectangular sheets after cross-linking. The sample was placed onto one surface of a pair of magnetically actuated grips and a displacement-controlled loading frame. Tension is applied via a servo motor in the $y$ direction, such that the displacement between grips are increased symmetrically; in this way, the center of the sample remain stationary in the laboratory frame of reference. A pre-crack is introduced along $x$ direction on an edge of the sample centered between the grips. During the experiment, the sample was immersed in water and illuminated with a dark-field configuration. A $10\times$ water-immersion objective (Nikon Plan Fluor $10\times$/0.30W) was mounted on microscope (Nikon eclipse Ti, tube lens magnification: $1.5\times$) to focus the scattered light from the particles onto the sensor of a high-resolution camera (Hamamatsu C13440, resolution: $2048\times2048$ pixels, bit depth: 16 bit), and imaged at up to 100 frames per second. As a consequence of the water-immersion objective and the water bath, the refractive index of the hydrogel (1.365) is well-matched to that of the surrounding water (1.333), and thus lensing effects near the crack tip were effectively eliminated, as can be seen from an sample image shown on the bottom right inset in Fig.~\ref{fig_J_Setup}(a). By carefully increasing mode~I loading, the initial crack starts to propagate quasi-statically, at a velocity of approximately \SI{10}{\micro\meter/\second}. At this rate of crack propagation, particles remain approximately stationary for the duration of the camera's exposure, eliminating motion blur. The microscope was adjusted to focus on the mid-plane of the sample, and a region of interest about \SI{1}{\milli\meter} ahead of the current crack tip was recorded by the camera at 3 frames per second.

The recorded images of crack, only if the crack travels in a straight path (ensuring pure mode~I loading), were first rotated to ensure that the crack travels horizontally and the loading is along vertical direction. A bandpass filter is then applied to all images to eliminate both image noise and large undesired structures appearing on images, e.g., floating dust.

\begin{figure}[!tb]
\centering
\includegraphics[width=0.95\textwidth]{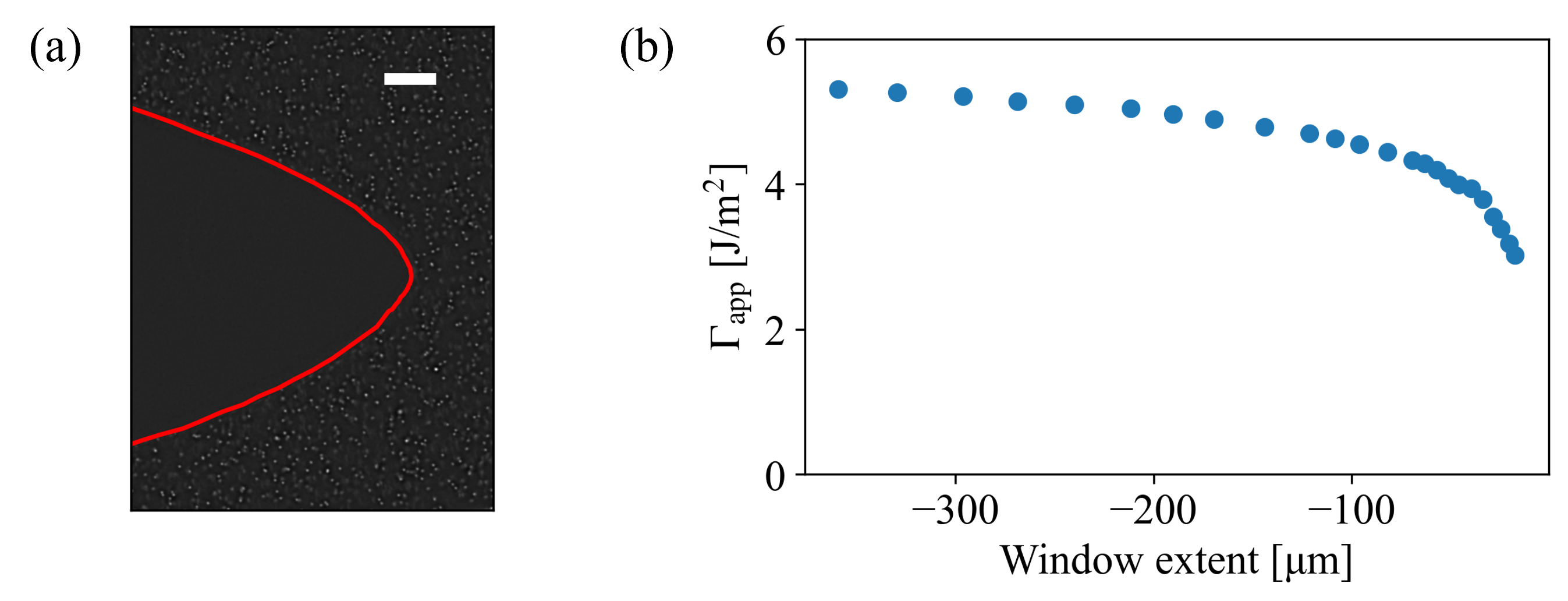}
\caption{CTOD measurement of a typical sample with particles. (a) Extracted CTOD from one of the typical samples with embedded particles used in the J-integral measurements. The CTOD is highlighted by the red curve. The scale bar is \SI{50}{\micro\meter}.(b) The apparent fracture energy $\Gamma_\mathrm{app}$ of the particle sample in (a) measured by the CTOD method. $\Gamma_\mathrm{app}$ asymptotes to $\Gamma^* = $\SI{1.94}{\joule / \meter \squared} at the crack tip, and converges to a background fracture energy as the window extent moves away from the crack tip. The curve is consistent as in the main text. This result shows that the CTOD measurement agrees with the J-integral measurements using particle tracking.}
\label{figS_w/t_particles}
\end{figure}

Naturally, one wonders whether the embedded particles used to track material displacements might alter the mechanical response, and furthermore the fracture mechanics, of the hydrogel. To evaluate whether this might be the case, we applied the CTOD analysis to a sample with embedded particles, and we find that neither the fracture energy, nor its functional form are altered in the particle embedded sample, as can be seen in Fig.~\ref{figS_w/t_particles}. This underscores our prior study of similar particle-embedded samples in the dynamic fracture context~\cite{Taureg2020Particles}. A plausible explanation for why the particles are truly passive, even in the context of fracture mechanics, is that they are significantly smaller than the typical yield scale in the material, where fracture processes occur; such scales have been evaluated in detailed studies of liquid-liquid phase separation in elastomer networks such as PDMA~\cite{style_phase}. 

To track the particles from the pre-processed images, an open source particle tracking algorithm, TrackPy~\cite{Trackpy,TrackPy1996}, is used. Two steps are necessary to successfully track particles - first, the particles must be located in each image, and second, the particles must be linked to their subsequent position in each succeeding image. In the first step, an estimated particle diameter is provided to construct a spatial filter; the particles identified by their intensity are then filtered by their mass (integrated brightness), percentile and eccentricity. To ensure sub-pixel resolution of particle positions, the uniformity of the distribution of the residual of each particles $x$ and $y$ coordinates is evaluated. For a typical particle image in our experiment, approximately 4000 particles are located with the field-of-view of \SI{800}{\micro\meter} by \SI{800}{\micro\meter}. 

The particles thus located are then linked between frames based on the particles' most recent velocity. Despite the large deformation and large rotation near the crack tip, we employ a sufficient frame rate to ensure the robustness of the velocity-based prediction of particle locations in subsequent frame. With accurate particle locations and reliable particle linking, the trajectories of particles are successfully tracked.

\begin{figure}[!tb]
\centering
\includegraphics[width=0.95\textwidth]{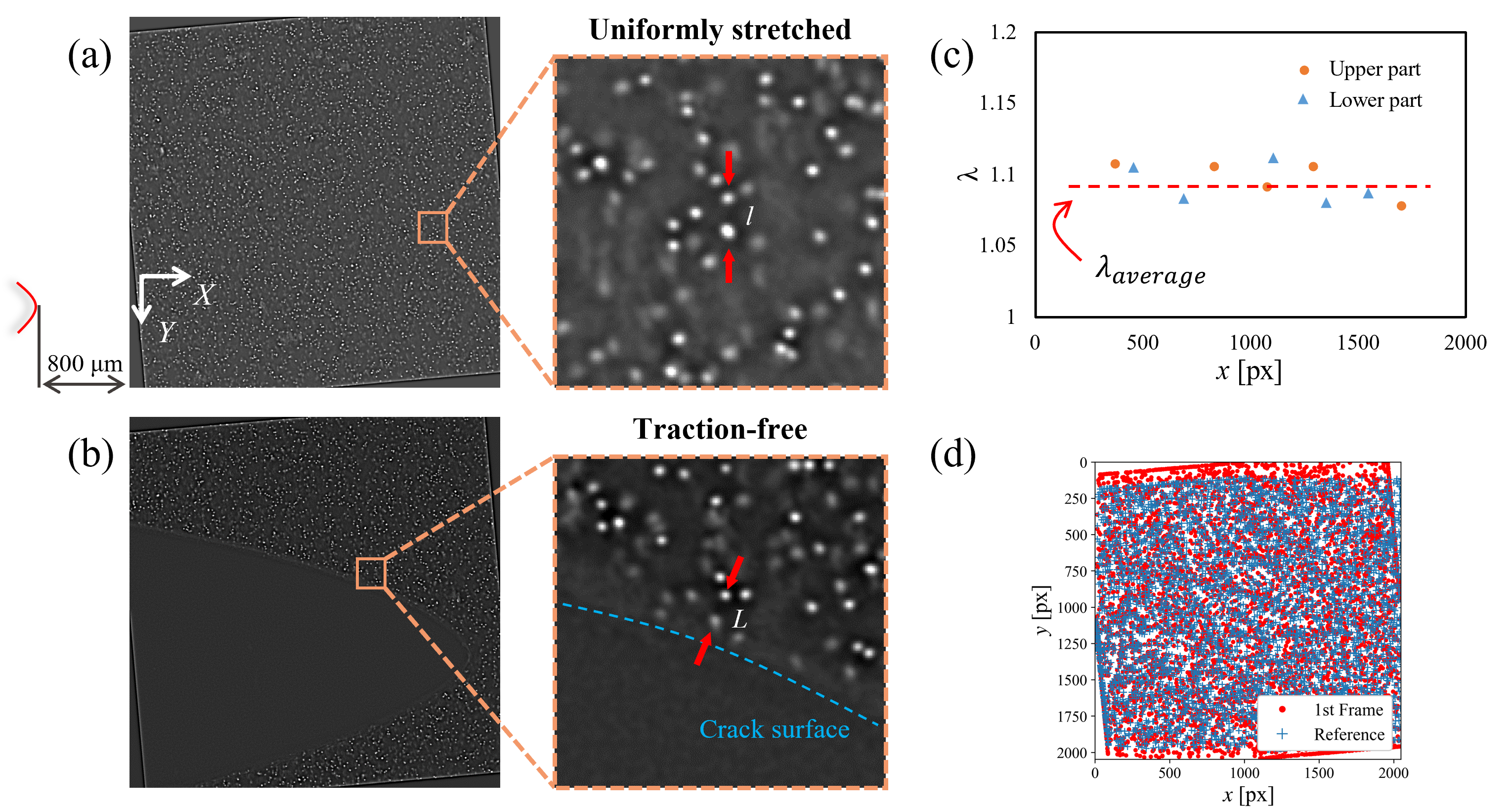}
\caption{Estimation of particles' location at undeformed state. The estimation is based on two frames of the recorded movie of crack propagation - the first frame (a) and the last frame (b). In the first frame, since the crack is still hundred of micrometers from the field-of-view (FOV), the hydrogel sample in the FOV is uniformly stretched. In the last frame, the crack has propagated through, and on the crack surface, the boundary is traction-free. 
For a pair of particles that is perpendicular to the crack surface in the last frame and vertically aligned in the first frame (indicated by red arrows in (a) and (b)), the distance between the two particles can be measured as $l$ in the first frame under uniform stretch and $L$ in the last frame near the traction-free boundary. By comparing the distance change of the particle pair, the uniform stretch in the first frame can be calculated as $\lambda = l/L$. (c) The uniform stretch is evaluated for 10 pairs of particles across the image from the left to the right, 5 on the upper crack surface and 5 on the lower crack surface. The average uniform stretch is measured as 1.09 with a standard deviation of 0.01. (d) The tracked particles (red dot) in the frame and their estimated locations at undeformed state (blue cross). With the measured uniform stretch $\Bar{\lambda}$ from (c) and the tracked particle location $(x,y)$ in the first frame, the particle locations at undeformed state can be retrieved as $(X,Y) = (x/(1-\nu(\Bar{\lambda}-1)),y/\Bar{\lambda}))$, with any arbitrary selection of the laboratory frame of reference, where $\nu$ is the Poisson's ratio and equals to 0.5 for our incompressible hydrogel. Note that this is only an example of one experiment. The same procedure is carried out on all three experiments and the measured uniform stretch values are all around 1.09.
}
\label{figS_X0}
\end{figure}

With the particle locations tracked at current states $\mathbf{x}$ and their locations at reference state $\mathbf{X}$ estimated as shown in Fig.~\ref{figS_X0}, displacement vectors for each particle can be directly calculated as $\mathbf{u} = \mathbf{x} - \mathbf{X}$. By performing finite difference on the displacement components of neighboring particles, the deformation gradient tensor $\mathbf{F}$ of each particle is estimated by
\begin{equation}
    \mathbf{F} = \mathbf{I} + \mathbf{\nabla_X u}.
\end{equation}
Then, it is straightforward to obtain the stretch and rotation tensor by polar decomposition, and thus to obtain the Lagrange finite strain tensor via the right Cauchy-Green deformation tensor. With an appropriate constitutive law, the displacements and deformation gradient tensor can be used to calculate the $J$-integral around the crack tip.

The $J$-integral is explicitly calculated from deformation fields by~\cite{rice_1968,knowles_1973,liu2020asymptotic}
\begin{equation}
    J=  \int_{c} \left( W n_1 - \sigma_{ij} n_{j} \frac{\partial u_i}{\partial x_1} \right)ds, i,j=1,2,
    \label{eqn_J}
\end{equation}
where $c$ is an arbitrary counterclockwise integral path with an outward unit vector $\mathbf{n}$, $W$ is the strain energy density, $\sigma_{ij}$ is the first Piola-Kirchhoff stress tensor, ${\partial u_i}/{\partial x_1}$ are the components of the displacement gradient tensor. To ensure the correct representation of energy release rate, $J$ must be calculated with strain energy density and stress tensor that are derived based on proper material model - in our case, the neo-Hookean material model.

The hydrogel used in our fracture experiments is known as an incompressible neo-Hookean material, whose strain energy density $W$ is defined as
\begin{equation}
    W = \frac{1}{2} \mu (I_1 - 3),
    \label{eqn_SED_3D}
\end{equation}
where $\mu$ is the shear modulus of the hydrogel, $I_1$ is the first invariable of the right Cauchy-green deformation tensor $\mathbf{C}$, i.e., $I_1=\text{tr}\mathbf{C}=\text{tr}(\mathbf{\Tilde{F}} ^ \text{T} \mathbf{\Tilde{F}})$. It it noted that the deformation gradient tensor $\mathbf{\Tilde{F}}$ used here is for 3D deformation; however, deformation gradient tensor $\mathbf{F}$ measured from our experiments is 2D. Considering the incompressibility of the hydrogel, i.e., $\det(\mathbf{\Tilde{F}})=1$, and our focal plane coincides with the mid-plane of the sample, Eq.~\ref{eqn_SED_3D} can be simplified as
\begin{equation}
    W = \frac{1}{2} \mu (\text{tr}(\mathbf{F} ^ \text{T} \mathbf{F}) + \det(\mathbf{F})^{-2} - 3),
    \label{eqn_SED_2D}
\end{equation}
where $W$ can be directly calculated from the measured $\mathbf{F}$. In this case, the in-plane Piola-Kirchoff stress is given by:
\begin{equation}
    \mathbf{\sigma} = \mu \mathbf{F} - \mu \det(\mathbf{F})^{-2}\mathbf{F}^{-\text{T}}.
    \label{eqn_Stress}
\end{equation}
Once we obtain the deformation gradient evaluated at each particle, we linearly interpolate the discrete data ($\mathbf{F}, W, \mathbf{\sigma}$) to construct the field data near the crack tip. Over these deformation fields, $J$-integrals are calculated according to Eq.~\ref{eqn_J} along rectangular integral paths.

\section{Results}

\subsection{Measurement of the apparent fracture energy in the near-tip region via CTOD}\label{subsection:CTOD}

LEFM predicts a parabolic CTOD for mode I cracks~\cite{anderson}, where the pre-factor of the parabola, $a$, is governed by the strain energy release rate $G_c$. By Griffith's energy balance criterion, $G_c$ is equal to the material fracture energy when the crack is critically loaded and propagating. In our experiments, the quasi-static propagation ensures that the crack is observed in a critical state, and thus satisfies energy balance. Under this condition, a parabola fitted to the measured CTOD (measured as described in section~\ref{subsection:mmctod}) can be used to directly compute the apparent fracture energy $\Gamma_{\text{app}}$ (See detailed derivation in section~\ref{section:mat-methods}). For a static mode~I crack, $\Gamma_{\text{app}}$ is given by 
\begin{equation}\label{eq_CTOD}
    \Gamma_{\text{app}} = G_c = \frac{3 \pi \mu}{8 a},
\end{equation}
where $\mu$ is the shear modulus of the material. The measured CTOD is perfectly parabolic in the region away from the crack tip, where the fracture behavior is known as \lq $K$-dominant\rq, as shown in Fig.~\ref{fig_CTOD_Setup}(c). In the small region near the crack tip, however, the CTOD deviates from the fitted parabola in a manner similar to $\delta_{dyn}$\cite{livne_breakdown_2008}, demanding further investigation. 

\begin{figure}[!tb]
\centering
\includegraphics[width=1.0\textwidth]{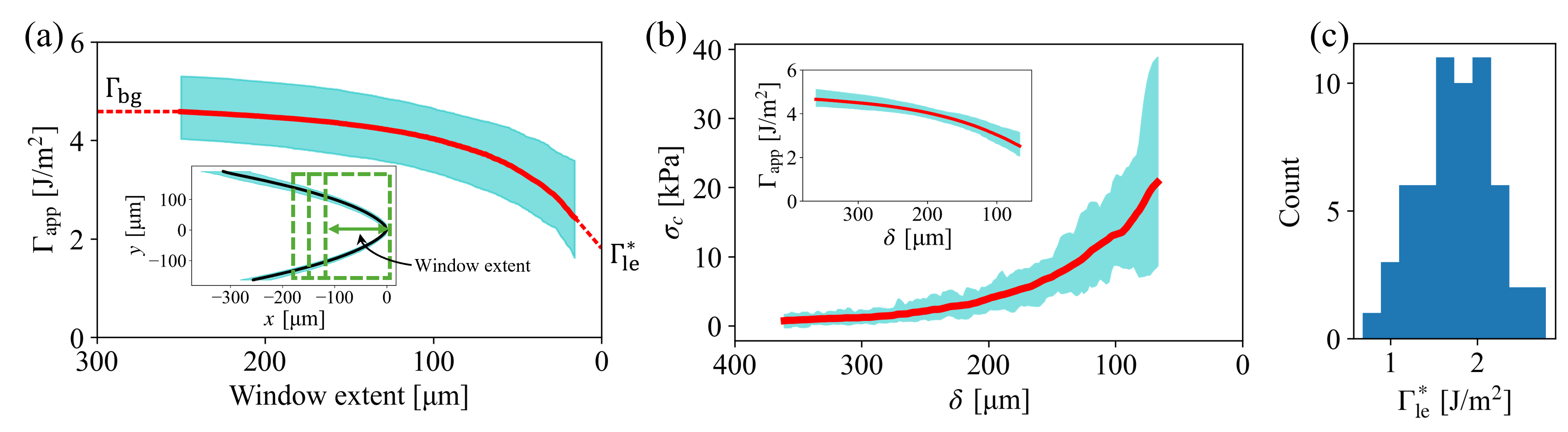}
\caption{CTOD evaluation of the apparent fracture energy near the crack tip. (a) Apparent fracture energy is plotted as a function of window extent for the parabolic fit. The mean and standard deviation are measured from 6 quasi-static crack samples consisting of 10 frames each. The mean is shown in the red curve and the standard deviation is indicated by the blue region for each value of the window extent. $\Gamma_\mathrm{{app}}$ approaches the background fracture energy $\Gamma_\mathrm{{bg}}$ as the window extent enters the K-dominant region. Using linear extrapolation of the data within \SI{25}{\micro \meter} distance to the crack tip, $\Gamma_\mathrm{{app}}$ asymptotes to the value of $\Gamma^*_\mathrm{le} = 1.77 \pm 0.44 \mathrm{J/m^2}$ as the window extent shrinks to zero at the crack tip. Inset: different window extents result in different values of $\Gamma_\mathrm{app}$ by altering the range of the data used for the fit. The left boundary of the window (green dashed rectangle) extends behind the crack, defining the window extent.
(b) The constitutive law of the cohesive zone is defined by the tensile stress as a function of the separation of the crack faces, $\sigma_{c} (\delta)$, for the brittle hydrogel network. $\sigma_{c}$ is determined from 
$\Gamma_\mathrm{{app}} (\delta)$, shown inset, as described in the main text.(c) Distribution of $\Gamma^*$ from the CTOD measurements, where $\Gamma^*$ is the asymptotic apparent fracture energy at the crack tip. The distribution is from 60 measurements obtained with 6 independent samples, which each frame treated individually. The mean is $1.77 \mathrm{J/m^2}$ and the standard deviation is $0.44 \mathrm{J/m^2}$.}
\label{fig_CTOD}
\end{figure}

$\Gamma_{\text{app}}$ is experimentally determined from the CTOD data by evaluating the best fit parabola over different window extents using a non-linear least squares routine and Eq.~\ref{eq_CTOD}. As the window extent changes for the evaluation of the best fit, the value of $a$ can change; thus, the value of $\Gamma_{\text{app}}$ is a function of window extent used to fit the parabolic geometry to the CTOD data. Analysis of the CTOD data in this manner shows that $\Gamma_{\text{app}}$ decreases near the crack tip, as shown in Fig.~\ref{fig_CTOD}(a). Further from the crack tip, $\Gamma_\text{{app}}$ plateaus to a background fracture energy $\Gamma_\text{bg}$. The variation of $\Gamma_{\text{app}}$ represents the loss of cohesion in the near-tip region. At the crack tip, $\Gamma_\mathrm{app}$ asymptotes to the value $\Gamma^*_\mathrm{le} = 1.77 \pm 0.44 \mathrm{J/m^2}$ by linear extrapolation of the data within a \SI{25}{\micro\meter} range near the crack tip (distribution of $\Gamma^*_\mathrm{le}$ for 60 measurements is shown in Fig.~\ref{fig_CTOD}(c)). $\Gamma^*_\mathrm{le}$ is the energy dissipation at the crack tip, and is distinct from the cohesive loss occurring within a ca.~\SI{100}{\micro\meter} neighborhood of the crack. These measurements suggest two distinct regions where energy is dissipated - a distributed damage region, and a highly-localized process region at the crack tip.

The constitutive behavior within the cohesive zone~\cite{freund} is defined by the variation of the cohesive stress $\sigma_c$ in this region; thus, one can evaluate $\sigma_c$ by the derivative of $\Gamma_\mathrm{app}$ with respect to the crack face separation distance $\delta$~\cite{freund}, as
\begin{equation}
     \sigma_{c}(\delta) = \frac{d\Gamma_\mathrm{app}(\delta)}{d \delta};
\end{equation}
a derivation of this expression presented in \textit{Materials and Methods}. The constitutive law of the cohesive zone is shown in Fig.~\ref{fig_CTOD}(b).

\subsection{Measurement of the \textit{J}-integral in the near-tip region via particle tracking} \label{subsection:J-integral}

The displacement vector field is determined using the microscopy method described in section~\ref{subsection:mmparticles}. Once the displacement vector field is determined via particle tracking for each particle, the deformation gradient tensor, $\mathbf{F}$, is estimated with finite-difference applied to the displacement components of the neighboring particles~\cite{Tyler_F_Estimator}. Using the estimate of the deformation gradient tensor, we calculate the strain energy density $W$ and the first Piola-Kirchoff stress tensor fields $\mathbf{\sigma}$ using the neo-Hookean material model. Expressions for $W$ and $\mathbf{\sigma}$ are provided in section~\ref{section:mat-methods}.

\begin{figure}[!th]
\centering
\includegraphics[width=0.99\textwidth]{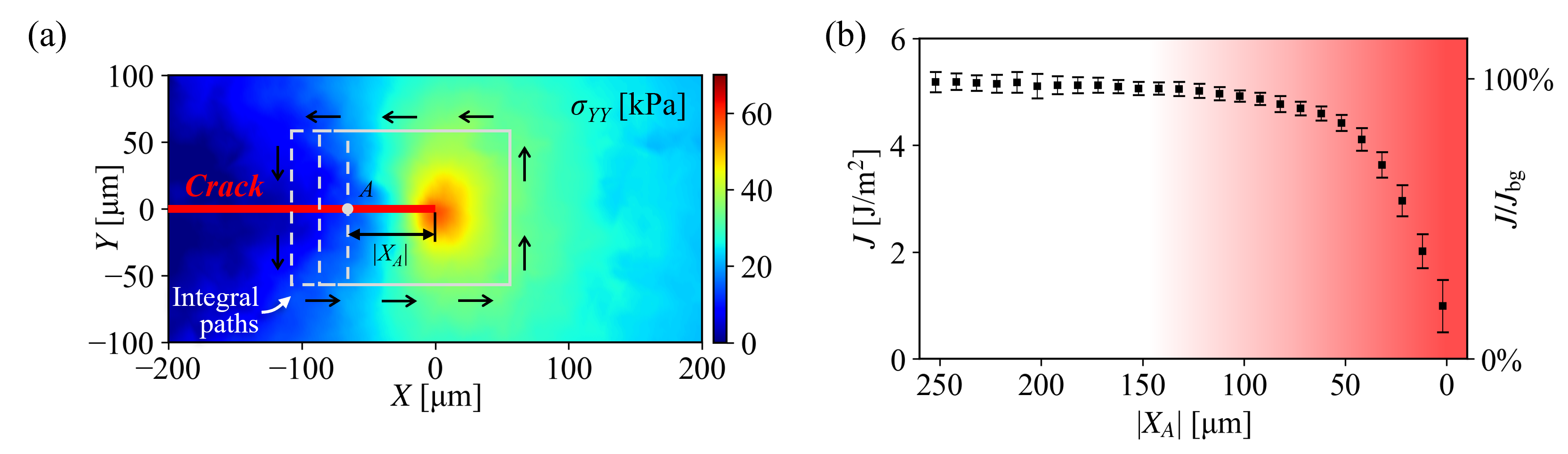}
\caption{$J$-integral evaluation with varying paths near the crack tip. 
(a) The stress concentration at the crack tip can be readily observed from the plot of the $\sigma_{YY}$ component of the first Piola-Kirchhoff stress, shown in the material frame of reference. The crack is indicated and labelled in red, extending from $(0,0)$ in the $-X$ direction. Several integral paths, used in the evaluation of the $J$-integral, are depicted by the dashed lines; only the left boundary location, indicated by the $x$ coordinate of its midpoint $A$, is varied upon changing the integral path. 
(b) The $J$-integral was evaluated for 13 separate measurements during crack propagation in 3 different gel samples. The left boundary is systematically varied in \SI{10}{\micro\meter} steps. Far from the crack tip, the change of integral window/contour does not affect the calculated value of $J$, consistent with its path-independence. When the window extent shrinks to less than \SI{150}{\micro\meter}, $J$ starts to drop, with a similar decreasing trend to what is observed from the CTOD measurement shown in Fig.~\ref{fig_CTOD}(a). The color gradient in the background represents the fractional deviation of the measured $J$ to a background $J_{\mathrm{bg}}$. The error bars indicate the standard deviation among the 13 measurements. 
}
\label{fig_J}
\end{figure}

Using these deformation fields, the $J$-integral is evaluated around the crack tip along a series of counterclockwise, rectangular paths as shown in Fig.~\ref{fig_J}(a). The left boundaries of the integral paths vary from the crack tip to the far crack tail (indicated by the $x$ coordinate of its midpoint $A$, $X_A$), while the other boundaries remain fixed at \SI{120}{\micro\meter} from the crack tip. Values of $J$ from integral paths with different left boundary locations are shown in Fig.~\ref{fig_J}(b). When the left boundary of the integral path is far away from the crack tip, i.e., $|X_A|>150\si{\micro\meter}$, the value of $J$ is nearly constant. In this region, the $J$-integral remains path-independent with a value of approximately \SI{5}{\joule/\meter^2} in agreement with the reported values of the fracture energy of this material~\cite{Wang2022JMPS}.

When $|X_A|<150\si{\micro\meter}$, a decrease of $J$ is observed, similar to the drop of $\Gamma_\mathrm{{app}}$ measured by the CTOD, as shown in Fig.~\ref{fig_CTOD}(a). We measured a background $J_\mathrm{{bg}}$ from a far-field (\SI{400}{\micro\meter} to the crack tip) integral path, and quantified the fractional decrease of the near-crack-tip $J$ to the background $J_\mathrm{{bg}}$. $J$ decreased by $\approx10$\% from $J_\mathrm{{bg}}$ for an integration window extending between \SI{150}{\micro\meter} and \SI{70}{\micro\meter} from the crack tip. A sharper decrease to approximately 20\% of $J_\mathrm{{bg}}$ is observed when $|X_A|<70\si{\micro\meter}$. This suggests, on the one hand, that a cohesive region with a finite size of approximately \SI{150}{\micro\meter} exists behind the crack tip; on the other hand, it also suggests that most of the energy dissipates in this cohesive region rather than ahead of the crack, despite the dominant stresses ahead of the crack tip.

\begin{figure}[tbh!]
\centering
\includegraphics[width=0.75\textwidth]{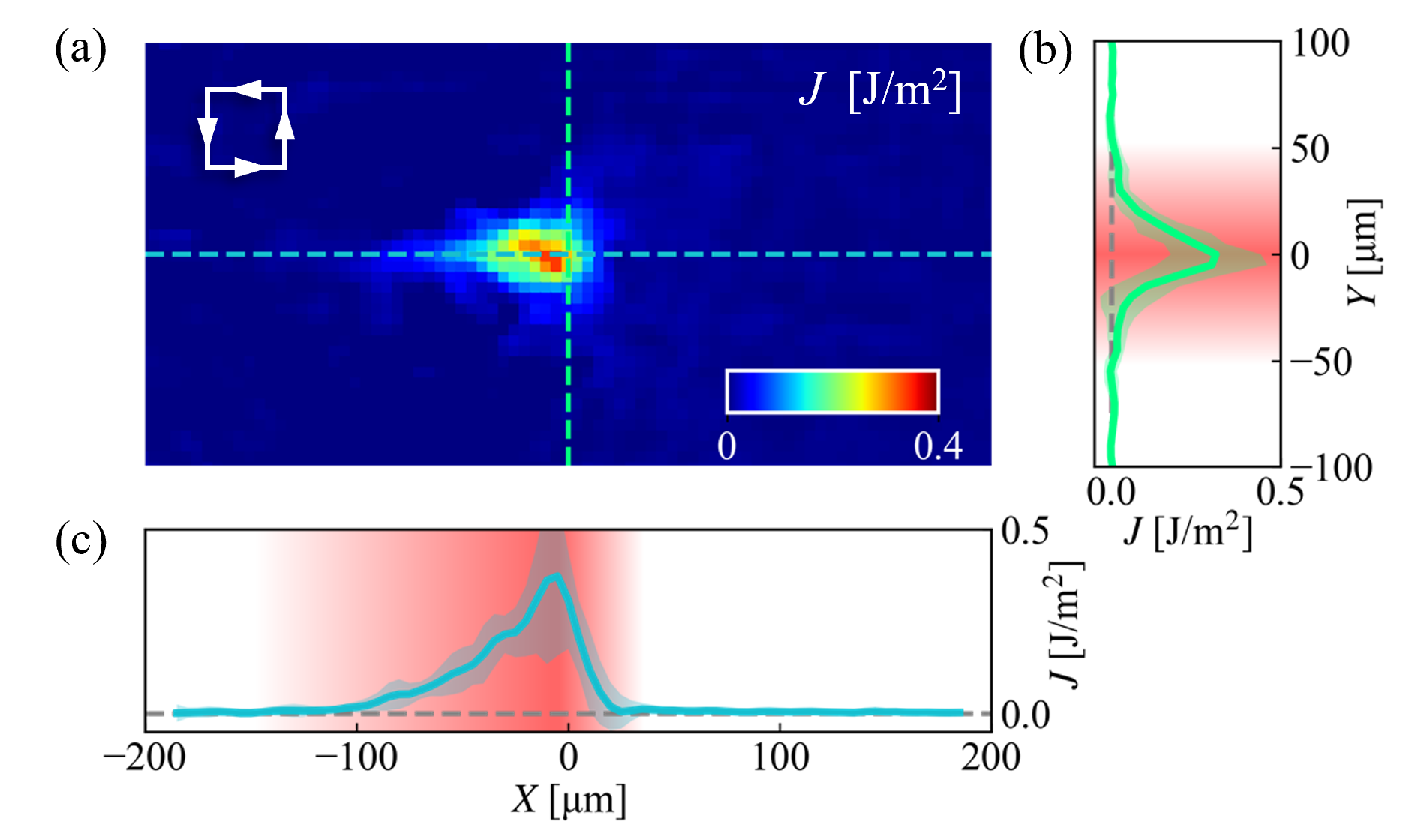}
\caption{Structure of the cohesive zone.
(a) The field of $J$ (averaged over 13 measurements) calculated locally by rastering the integral path which is defined by the perimeter of small square windows (top left) with side length of \SI{10}{\micro\meter}. 
Far from the crack tip, $J$ is approximately zero, suggesting that no energy is dissipated in these regions; however, close to the crack tip, $J$ increases, quantitatively indicating the energy dissipated locally within the window through the material decohesion.
To evaluate the structure of this cohesive zone, the values of $J$ along $X=0$ (green dashed line) and $Y=0$ (blue dashed line) are extracted and presented in (b) and (c), respectively. $X$ and $Y$ stand for the coordinates of the center of the window.
(b) Localized J values increase near the crack tip and are symmetric about the crack along $Y$. The energy dissipation is localized in a \SI{50}{\micro\meter} region on either side of the crack. 
(c) Along $X$, the localized values of $J$ are not symmetric about the crack tip; Energy dissipation is observed only in a very limited area ahead of the crack tip (\SI{20}{\micro\meter}) but extends approximately \SI{150}{\micro\meter} into the crack tail, consistent with our observation from the CTOD measurement shown in Fig.~\ref{fig_CTOD}(a) and the windowed $J$ evaluation in Fig.~\ref{fig_J}(b).
}
\label{fig_J_windowing}
\end{figure}

To probe the local structure of energy dissipation, we evaluate the $J$-integral along the perimeter of \SI{100}{\micro \meter \squared} square windows. A map of the dissipation emerges as these windows are rastered around the crack tip with a step of \SI{5}{\micro\meter}. This process is repeated for 13 individual recordings of the crack tip field, which are then averaged as shown Fig.~\ref{fig_J_windowing}(a). We exploit the path independence of $J$ by recognizing that any finite value returned from the evaluation of $J$ represents the local dissipation in the square window region. The local energy dissipation is primarily distributed along $-X$, where the material has just fractured, as shown in Fig.~\ref{fig_J_windowing}. Away from the crack tip, the localized value of $J$ decreases to zero, indicating that no dissipation takes place in these regions, as is expected for $K$-dominance.

The structure of the cohesive zone can be characterized by taking slices of the $J$ map along the crack path $Y=0$ and perpendicular to it along $X=0$. The energy dissipation represented by the value of $J$ is symmetric and localized in a \SI{50}{\micro\meter} region on either side of the crack face, as shown in Fig.~\ref{fig_J_windowing}(b). In the direction of crack advancement, the value of $J$ is non-zero only \SI{20}{\micro\meter} ahead of the crack tip, but extends approximately \SI{150}{\micro\meter} behind the crack, as can be seen in Fig.~\ref{fig_J_windowing}(c). This asymmetry of energy dissipation is consistent with our observation from CTOD measurement shown in Fig.~\ref{fig_CTOD}(a) and the windowed $J$ evaluation in Fig.~\ref{fig_J}(b). From the color-plot of $J$ in Fig.~\ref{fig_J_windowing}(a), the cohesive zone for a quasi-static crack in this hydrogel looks like a droplet moving along a surface in the direction of the crack with dimensions of the order of \SI{100}{\micro \m} along each axis.

\subsection{Evaluation of strain stiffening}\label{subsection:strainstiffening}

Among the various mechanisms that might alter the constitutive behavior under large stretch near the crack tip, strain stiffening is the primary means by which the observed sharper CTOD could emerge~\cite{Long_Hui_2015}. In order to evaluate whether strain stiffening occurs for the hydrogel material used in our experiments, we evaluate the constitutive response at the largest values of stretch observed near the crack tip with the particle tracking measurements. 

\begin{figure}[!tb]
\centering
\includegraphics[width=0.95\textwidth]{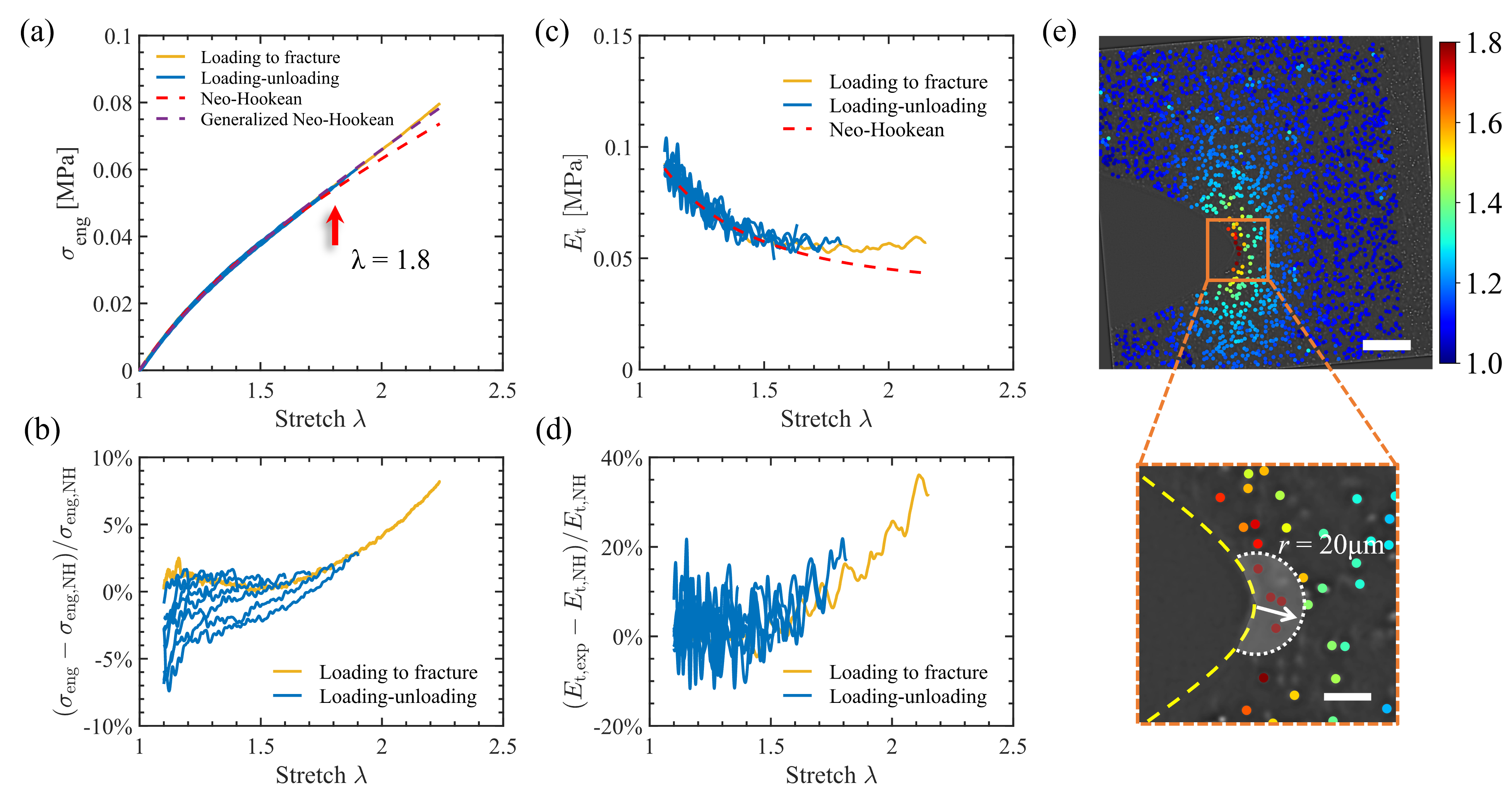}
\caption{(a) Stress-stretch ($\sigma_\mathrm{eng}-\lambda$) curves of two uniaxial tensile tests of intact hydrogel samples - a monotonic test until sample fracture ($\lambda_\mathrm{max} \approx 2.24$), and a cyclic test with stretch range from 1 to 1.9. The experimental curves are fitted to the incompressible neo-Hookean model and the generalized neo-Hookean model~\cite{Long_Hui_2015}, as shown in the red and purple dashed curves respectively. The experimental data starts to visibly deviate from the theoretical prediction at stretches larger than 1.8. (b) The percentile deviation of the experimentally measured stress to the neo-Hookean prediction. The data with $\lambda < 1.1$ is truncated due to the huge numerical fluctuation of the deviation value as the stress approach zero at $\lambda = 1$. (c) The tangent moduli ($E_\mathrm{t}$) for both experimental data and neo-Hookean model. (d) The percentile deviation of tangent moduli measured from experiments to neo-Hookean model. (e) The maximum principle stretch field near the crack tip measured by particle tracking. The stretch remains slightly larger than 1 until very close to the crack tip. Inset: a magnified view of the stretch field in the immediate vicinity of the crack tip (crack indicated by yellow dashed line). The region with stretch larger than 1.8 is roughly characterized by a circle with radius of \SI{20}{\micro\meter}. The scale bar is \SI{100}{\micro\meter} in the main plot and \SI{20}{\micro\meter} in the inset.}
\label{figS_UniStretch}
\end{figure}

To characterize the mechanical response of our hydrogel material, we load it in uniaxial tension and evaluate the stress as a function of stretch, as shown in Fig.~\ref{figS_UniStretch}(a). The $\sigma_\mathrm{eng}-\lambda$ curves are fitted to both the incompressible neo-Hookean model and the generalized neo-Hookean model respectively\footnote{In the generalized neo-Hookean material model, the strain energy density function is given by $W = \frac{\mu}{2b} \left[ \left( 1+ \frac{b}{n} (I-3)\right)^n - 1\right]$, where $\mu$ is the small-strain shear modulus, and $b$ and $n$ are material parameters~\cite{geubelle1994finite}. }. The generalized neo-Hookean model can account for strain stiffening and material yield through two parameters $b$ and $n$, where $b$ is a parameter corresponding to yield and $n$ is a stiffening exponent~\cite{geubelle1994finite}. When $n = 1$, the generalized neo-Hookean model reduces to the neo-Hookean model with the strain energy density given by $W = \frac{\mu}{2}\left( I - 3 \right)$. A fit to the generalized neo-Hookean model can be sensitive, particularly for data that are well-described by the neo-Hookean material model; nevertheless, using non-linear least squares we obtain a good fit with $n = 1.07 \pm 0.02$. Notably, this value is well below the threshold above which the strain stiffening leads to a sharper CTOD at $n = 3/2$~\cite{Long_Hui_2015}. 

Mild strain stiffening emerges in comparison to the classical neo-Hookean material response; we evaluate the relative difference between our material and the neo-Hookean response via the relative percentage difference of the stress, the tangent modulus and the relative percentage difference of the tangent modulus, as shown in Fig.~\ref{figS_UniStretch}(b)-(d). Under small stretches $\lambda$ and even moderate stretches ($\lambda < 2$), the deviation generally remains within $5\%$. The tangent modulus $E_\mathrm{t}$ decreases with the incremental stretch. For $\lambda < 1.6$, the theoretical tangent modulus, $E_\mathrm{t,NH}$, agrees well with the experimental tangent modulus, $E_\mathrm{t,exp}$. For a larger stretch, $E_\mathrm{t,exp}$ starts to deviate from the neo-Hookean prediction $E_\mathrm{t,NH}$ and shows a larger modulus, implying that the hydrogel is stiffer than neo-Hookean prediction in this small region. However, the region where stretch is larger than 1.8 is only within \SI{20}{\micro\meter} around the crack tip, as shown in Fig.~\ref{figS_UniStretch}(e).  Since our observation of apparent fracture energy drop and J-integral decrease has a characteristic length of around \SI{100}{\micro\meter}, strain stiffening is excluded from the possible reasons for the energy and $J$ drop. We then conclude that the energy drop is a result of energy dissipation within the cohesive zone.






\section{Discussion and Conclusions}\label{section:discussion}

Our micro-scale experiments demonstrate how a canonical brittle material converts strain energy into both distributed, and much more localized, dissipation at a crack tip. Within this material, there is an enormous separation of scales - the typical scale of the sample is centimetric, whereas the near-tip dissipation region extends no further than a couple of hundred microns away from the very tip of the crack. It is interesting to note that the CTOD data show two regions where dissipation occurs - a finite-sized cohesive region that extends approximately \SI{100}{\micro\meter} from the crack tip and an optically unresolved region, in which the remainder of the strain energy, $\Gamma^*_\mathrm{le}$, is dissipated. The cohesive region reveals the delocalized energy dissipation due to the distributed chain scission in the hydrogel network. The size of the cohesive region is consistent with the molecular damage zone directly visualized by fluorescence mechanophores in elastomer~\cite{slootman2020mechanophore}, and our observation confirms the imperfection of polymer networks in the Irwin-Orowan model~\cite{suo_1}. We discuss alternative mechanisms for the observed CTOD data, and provide numerical simulations for these mechanisms, in Appendix~\ref{appendix:CTOD}.

Having eliminated poro-elastic stress relaxation and strain stiffening from the possible mechanisms responsible for modifying the stress state at the crack tip, we are left with cohesive loss as the primary remaining mechanism that can explain our observations. In this case, we ought to have a quantitative estimate of the cohesive zone scale that should agree with the observed scale of the measured cohesive region. Indeed, we can recover an estimate for this scale by comparing the measured fracture energy to the highly repeatable work of fracture (integrated work on the material prior to failure in a defect-insensitive sample)~\cite{suo_1, long2021fracture}. Here the fracture energy is $\Gamma \approx $ \SI{5}{\joule \per \meter \squared} and the work of fracture is calculated by numerically integrating the loading curve as $\mathcal{W} = $ \SI{54}{\kilo \joule \per \meter \cubed}; thus, the cohesive lengthscale is $\ell_{coh} \approx$ \SI{93}{\micro \meter}, a value in almost perfect agreement with the scale of the cohesive zone we measure, identifying the importance of cohesive loss as the primary mechanism responsible for determining the state of stress at the crack tip in this brittle hydrogel. Notably, similar analysis yields a similar scale for the cohesive zone in brittle high-tensile steels~\cite{Pardoen}, highlighting the universality of this analysis for different materials.

How might our dissipation scale and sharp feature in the cohesive region near the crack tip relate to the length scale $\delta_{dyn}$ observed in extensive experiments on dynamic fracture~\cite{livne_breakdown_2008, livne_2010, goldman_acquisition_2010, goldman_boue_failing_2015}? This point is less clear, as that feature has long been ascribed to the $1/r$ scaling predictions of prior theoretical works~\cite{knowles_1973, bouchbinder_1/r_2009}. However, it is known that this scale grows with crack speed~\cite{livne_breakdown_2008, goldman_acquisition_2010}, as does the fracture energy~\cite{goldman_acquisition_2010}; thus, if $\delta_{dyn}$ somehow relates to the scale of the cohesive zone, it might account for the velocity dependence of the fracture energy observed in dynamic fracture; however, we cannot rule out strain stiffening at the strain rates realized in dynamic fracture. Direct evaluation of the material constitutive response must be evaluated at the rates typical of dynamic fracture to determine whether the critical 3/2 strain stiffening exponent of the generalized neo-Hookean model is exceeded at these strain rates; given the amplitude of the stretch realized near the crack tip, this is a significant experimental challenge.

The chemistry of the polyacrylamide hydrogel used in this study was selected for its brittle material response; however, polyacrylamide gels, among other hydrogels, have an incredible spectrum of mechanical responses~\cite{cohen_1992, suo_1, suo_2, suo_3, suo_4, suo_5, suo_6, zhao_chemrev}. It is likely that hydrogels with a visco- or poro-elastic constitutive response might have a different cohesive zone structure. In this work, we employ a general approach for directly measuring the cohesive zone law, and quantitatively evaluate effects such as strain stiffening and solvent transport. When a material has an alternative dominant dissipation mechanism through, e.g., plasticity as in the case of double-network hydrogels~\cite{gong, kolvin_how_2018}, this method must be considered carefully. Indeed, cracks in brittle materials and cracks in plastic materials have different crack tip fields~\cite{knowles_1973, anderson, hutchinson, rice_rosengren}.

Interestingly, our measurement of the apparent fracture energy identifies a baseline value of the fracture energy \emph{within} the process zone, which we call $\Gamma^{*}_\mathrm{le}$; this is the energy required to cause the rupture of bonds at the crack tip, and there is a separation of scales between this process and the cohesive zone. This quantity can be estimated from the mass fraction of the monomer and the bond energy using a characteristic molecular scale of the order of \SI{2}{\angstrom} - for polymerized acrylamide, the C-C bond energy is about \SI{7e-19}{\joule}; the cross sectional area of the molecule is approximately \SI{4e-20}{\meter}$^{2}$, and the mass ratio is about 10\%, yielding an estimated \SI{1.7}{\joule \per \meter \squared}, in rough agreement with the measured value of $\Gamma^{*}_\mathrm{le}$. Notably, the CTOD measurements and the $J$-integral measurements indicate different values for $\Gamma^{*}$. We attribute this to inaccuracies arising from linear extrapolation of $\Gamma_\mathrm{app}$. The linear extrapolation likely overestimates the value of $\Gamma^{*}$, as the \emph{curvature} of $\Gamma_\mathrm{app}$ is negative as the window extent asymptotically vanishes.

We have evaluated the energy loss occurring within the cohesive region of a brittle hydrogel. This material is well-described by a neo-Hookean constitutive law over all values of the stretches measured here, with negligible strain stiffening. Poroelastic modifications to the stress might occur, albeit at scales significantly smaller than the observed cohesive zone, at the limit of the spatial resolution of our measurements. The loss of cohesion will likely vary with different materials. Our measurements can motivate the use, and the spatial structure, of the phase-field in numerical calculation of brittle fracture~\cite{deLorenzis_2015}, even in dynamic, unstable cases~\cite{Bleyer_2017}. Notably, our observations confirm the long-standing idea of scale separation between the material's elastic response and the small scale region at the tip of a brittle crack - the so-called autonomy of the crack tip\cite{rice_1968}, as there are 4 orders of magnitude separating the sample width and the extent of the cohesive zone, ensuring that effects from the boundaries of the sample in the crack plane do not alter the structure of the cohesive zone. In this work, we have characterized and measured the local dissipation during failure of a brittle solid using high-precision, high magnification measurements of the kinematics of near-crack-tip fields, opening a door toward future studies and extensions of this approach to other material systems.

\appendix

\section{Alternative Constitutive Response at the Crack Tip}
\label{appendix:CTOD}

While it is impossible to determine the constitutive response of the material near the crack tip from purely kinematic data, we can use estimates of different possible mechanisms to evaluate whether they play a dominant role in altering crack tip stress. One primary mechanism at play in hydrogels is that of stress relaxation due to solvent transport. A typical lengthscale used to evaluate the scale over which pore fluid migrates to relieve stress is derived from the ratio of crack speed, $v$, to the solvent diffusion constant, $D$~\cite{Baumberger_2020}; this can readily be compared to the scale of the cohesive zone measured in our work. For the cracks in our hydrogel, $v = $ \SI{10}{\micro \meter \per \second} and $D = $ \SI{1e-10}{\meter \squared \per \second}\cite{kalcioglu2012macro}; thus the length scale over which stress is relaxed by solvent migration is less than \SI{10}{\micro \meter}; this is significantly smaller than the scale of the cohesive region we have identified. 

An alternative mechanism that might be at play, and has indeed been invoked to explain dynamic fracture response~\cite{bouchbinder_1/r_2009, goldman_boue_failing_2015}, is that of strain stiffening in the context of a generalized neo-Hookean constitutive response. Indeed, it is known that below a certain threshold of strain stiffening, neo-Hookean solids have a parabolic CTOD, and that a strain stiffening parameter exceeding 3/2~\cite{Long_Hui_2015} must be present to observe the characteristic $1/r$ scaling\cite{knowles_1973}. For our material, we directly measured the material response subject to uniaxial tension. We find that the hydrogel we use is nearly ideally neo-Hookean at the strain rates we tested (and those realized in our fracture experiments); thus, we anticipate a parabolic CTOD, and any `sharpness’ in our measured CTOD is due to an alternative mechanism rather than strain stiffening of a generalized neo-Hookean material model. 

\begin{figure}[!htb]
\centering
\includegraphics[width=0.95\textwidth]{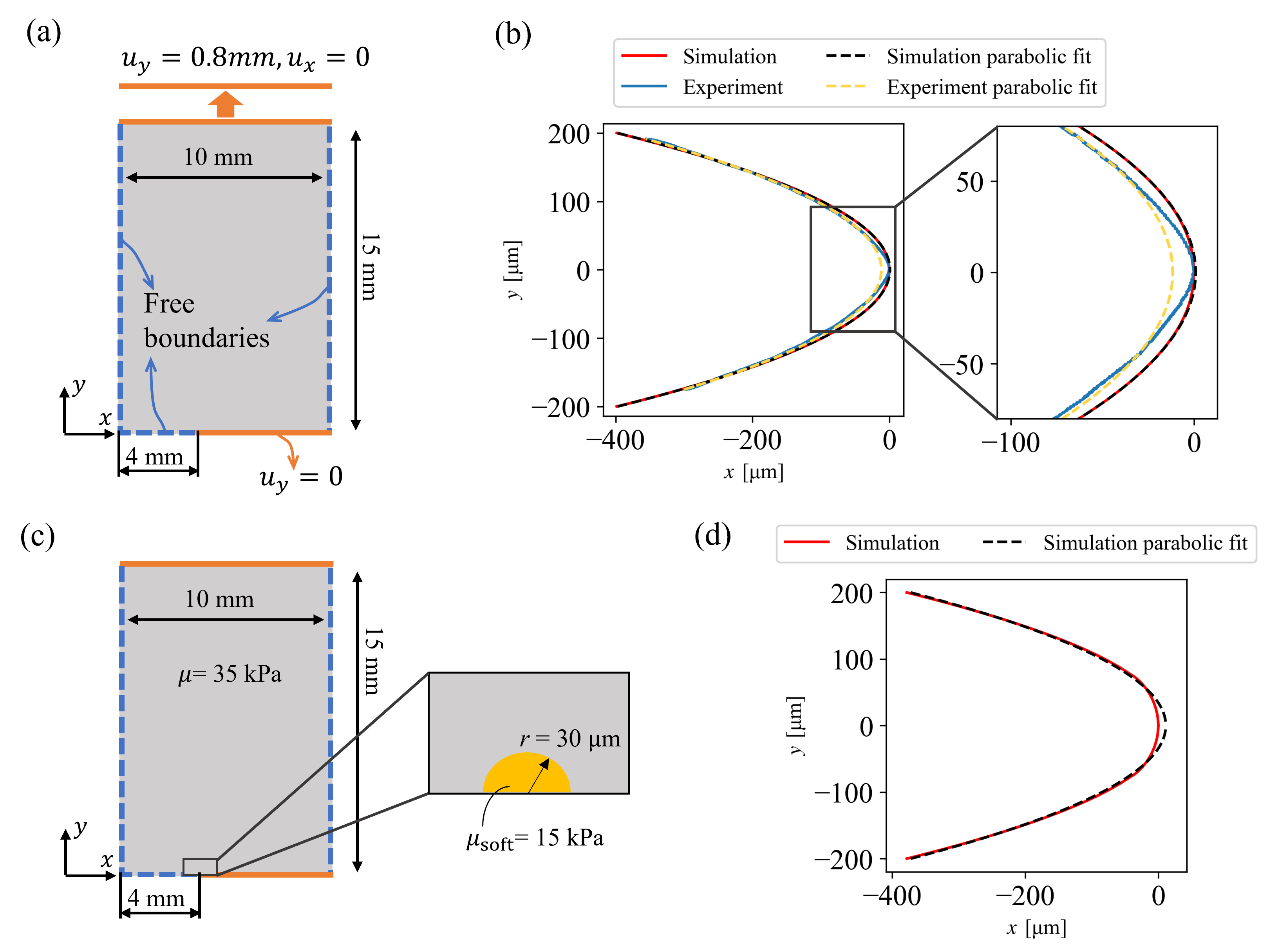}
\caption{Comparison of experimental and numerical CTODs from finite element simulation of half the sample and their parabolic fittings. (a) 2D finite element simulation setup for plane stress. The \SI{10}{mm} $\times$ \SI{15}{mm} rectangular geometry takes half the size of the experimental sample by symmetry. The displacement conditions to simulate the crack under mode~I loading are shown on each boundary. The material uses the neo-Hookean constitutive model with the shear modulus of \SI{35}{kPa}. The mesh is built with 166033 vertices after adaptive mesh refinement. (b) CTOD and parabolic fittings from simulation and experiment. The upper half of of the CTOD from the simulation is mirrored over the $x$ axis and plotted in red solid curve. An exemplary experimental data is plotted in blue solid curve. The crack tip of both CTODs are moved the origin. The parabolic fitting of the CTODs are shown in black and yellow dashed curves respectively. The right part is magnified from the area in the rectangular box from the left. (c) 2D finite element simulation setup for gel sample with a weakening region around the crack tip. The semicircle highlighted in yellow at the crack tip has the radius of \SI{30}{\micro \meter}. The shear modulus in this region is $\mu_\mathrm{soft} = \SI{15}{kPa}$, while the rest of the material remains $\mu = \SI{35}{kPa}$. The dimensions and the boundaries conditions are the same as in (a). (d) CTOD from simulation (red solid curve) in (c) and the parabolic fitting (black dashed curve).}
\label{figS_numerical}
\end{figure}

This is confirmed with numerical calculations carried out with COMSOL Multiphysics 6.0, where the CTOD remains parabolic to within the grid resolution (ca. \SI{1}{\micro \meter} to scale) for an ideally neo-Hookean solid, as shown in Fig.~\ref{figS_numerical}(b). Further calculations show that a region with lesser modulus leads to a \emph{blunter} CTOD, as shown in Fig.~\ref{figS_numerical}(d).

\clearpage


 \bibliographystyle{elsarticle-num} 
 \bibliography{cas-refs}





\end{document}